

\magnification=1200
\hsize=6.truein
\vsize=8.5truein   
\hoffset=.4truecm

\baselineskip=13pt plus 1pt minus 1pt
\parskip=0pt
\parindent=1.5pc

\def\mydate{August 25, 1994}

\def\normal{\baselineskip=13pt plus 1pt minus 0pt}

 1
 1
 1


\font\eightrm=cmr8
\font\ninerm=cmr9

\font\nineit=cmti9
\font\eightit=cmti8

\font\ninebf=cmbx9

\def\myref#1{$^{#1}$}

\def\mysec#1{\vskip .6cm  \leftline{\ninebf #1} \vglue .2cm}

\def\short#1{\hbox{$\kern .1em {#1} \kern .1em$}}

\def\big{\displaystyle \strut }

\def\N{\kappa}

\def\vp{ {\vec p} }

\def\L{ {\cal L} }

\def\E{ {\cal E} }
\def\O{{\rm O}}

\def\ep{\epsilon}
\def\eps{\varepsilon^{\mu\nu\rho}}
\def\d{\partial}

\def\tr{ {\,{\rm tr}\,} }
\def\Tr{ {\,{\rm Tr}\,} }

\def\d{\partial}
\def\la{\raise.16ex\hbox{$\langle$} \, }
\def\ra{\, \raise.16ex\hbox{$\rangle$} }
\def\st{\, \raise.16ex\hbox{$|$} \, }
\def\go{\rightarrow}

\def\eff{ {\rm eff} }

\def\Spect{ {\cal S}\, }
\def\pert{ {\rm p.v.} }
\def\gr{ {\rm g.s.} }

\def\slave{ {l_{\rm ave}^{\,2} } }
\def\ilave{ {l_{\rm ave}^{-1}} }

\def\psibar{ \psi \kern-.65em\raise.6em\hbox{$-$} }
\def\Lbar{ {\cal L} \kern-.65em\raise.6em\hbox{$-$} }
\def\Dmu{ D^\mu \kern-1.4em\raise.7em\hbox{$\leftarrow$} ~ }
\def\Dnu{ D^\nu \kern-1.4em\raise.7em\hbox{$\leftarrow$} ~ }

\def\smallhbar{{\eightit h} \kern-.8em\hbox{\eightrm\char'26}}
\def\smallnoteq{{\eightrm =} \kern-0.97em\hbox{\eightrm\char'57}$\,$}

\rm

\baselineskip=9pt
\line{\eightrm Preprint from \hfil   UMN-TH-1308/94}
\line{\eightrm University of Minnesota \hfil \mydate}
\vglue .6cm

\baselineskip=15pt

\centerline{\ninebf SPONTANEOUS BREAKDOWN OF THE LORENTZ INVARIANCE}
\centerline{\ninebf IN THREE-DIMENSIONAL GAUGE
THEORIES\footnote{$^\dagger$}{\nineit To appear in the Proceedings of
PASCOS '94.}}


\vskip .8cm

\baselineskip=10pt

\centerline{\eightrm  YUTAKA HOSOTANI}
\centerline{\eightit School of Physics and Astronomy, University of Minnesota,
               Minneapolis, MN 55455}

\vskip .8cm

\centerline{\eightrm ABSTRACT}
\vskip .2cm


{\eightrm \baselineskip=10pt
\rightskip=3pc \leftskip=3pc \noindent
In a class of renormalizable three-dimensional abelian gauge theory the Lorentz
invariance is spontaneously broken by dynamical generation of a magnetic
field {\eightit B}.    The true ground state resembles that of the quantum
Hall effect.   An originally topologically massive photon becomes gapless,
fulfilling the role of the Nambu-Goldstone boson associated with the
spontaneous breaking of  the Lorentz invariance.  We give a  simple
explanation  and a  sufficient  condition for  the spontaneous breaking of the
Lorentz invariance  with the aid of the Nambu-Goldstone theorem.
The decrease of the energy density by {\eightit B}{\smallnoteq}0 is understood
mostly due to the shift in zero-point energy of photons.
\smallskip}


\normal

\mysec{1. Variational Ground State}
In a wide class of three-dimensional gauge theories described by\myref{1,2}
$$\eqalign{
\L = &- {1\over 4} \, F_{\mu\nu}F^{\mu\nu} - {\N_0 \over 2} \,
\eps A_\mu \d_\nu A_\rho   \cr
&\hskip .0cm + \sum_a  {1\over 2} \, \big[ \,\psibar_a \, , \,
  \big( \gamma^\mu_a (i \d_\mu + q_a A_\mu)
    - m_a \big) \psi_a \, \big] ~, \cr}  \eqno(1) $$
the Lorentz invariance is spontaneously broken by dynamical generation of
a magnetic field $B$.  We have constructed a variational ground state
which has $B \short{\not=}0$ and has a lower energy density than the
perturbative vacuum.  The theory is renormalizable.

Taking advantage of charge-conjugation invariance of (1), one can take $q_a
\short{>}0$ without loss of generality.  Chirality of a Dirac particle is
defined by
$\eta_a \short{=} {i\over 2} \, \Tr \gamma_a^0\gamma_a^1\gamma_a^2$
$\short{=} \pm 1$ (for $m_a \short{\ge}0$).  It is equivalent to the sign of
$m_a$ for
$m_a \short{\not=}0$ with $\eta_a\short{=}+$, as can be understood by making a
transformation $\{ \gamma^\mu \} \go \{ -\gamma^\mu \}$.

Suppose that a uniform $B \short{\not=}0$ is dynamically generated.  The
energy spectrum of Dirac particles defines Landau levels:
$$\eqalign{
 E_0 ~  &= \hskip .5cm   \ep (\eta_a B) \cdot m_a  \hskip 1.5cm (n=0) \cr
 E_n^\pm  &= \pm  ( m_a^2 + 2nq_a|B|)^{1/2} \qquad (n \ge 1). \cr}
    \eqno(2)  $$
Each Landau level has the density of states $q_a |B|/2\pi$.
There is asymmetry in the $n\short{=}0$ modes (zero modes).  They exist in
either positive or negative energy states, or in other words, only for either
particles or anti-particles.   In the $m_a \short{\go} 0$ limit
the lowest energy level
$|E_0|$  approaches zero so that the ground state energy does
not depend, at least in the leading order in perturbation theory,  on whether
the lowest level is partially filled or not.  In other words the perturbative
ground states in the $m_a \short{\go} 0$ limit are infinitely degenerate.  The
degeneracy is lifted by quantum corrections.

We consider variational ground states in which these lowest Landau levels
are either empty or completely filled.\myref{3}   Accordingly a filling
factor $\nu_a\short{=} 0$ or $1$ is assigned to each species of fermions.  A
variational ground state is denoted by $\Psi_\gr (B, \{ \nu_a \} )$.  In
general one should consider more general filling factors.   Among
them Laughlin-type states with $\nu\short{=} {1\over 3}$, ${1\over 5} \cdots$,
are good candidates.  In this article we confine ourselves to
$\nu \short{=}0$ or 1.  We shall show that when $\{ \nu_a \}$
satisfies a certain condition, the state $\Psi_\gr (B, \{ \nu_a \} )$ has a
lower energy density than the perturbative vacuum with $B \short{=} 0$.

\mysec{2. Consistency Condition}

With fixed Dirac matrices   one can continuously change the value
of fermion mass $m_a$ from positive to negative.  Except for zero-modes
 in the spectrum  (2) all positive (negative) frequencies remain
positive (negative). However, there results a crossing in zero modes.
Positive frequency zero-modes  become negative frequency zero-modes, or vice
versa.  Its implication is that under the continuous change of $m_a$
an empty state $\nu_a=0$ with positive $m_a$ becomes a completely filled
state $\nu_a=1$ with negative $m_a$.

To see it, let us suppose $\eta_a B \short{>}0$.  For a positive $m_a$
zero modes exist only for positive energy solutions.  Hence we have
schematically
$$\psi = \sum_{k} \Big\{ \sum_{n=0}^\infty a_{nk} u_{nk}(x)
  + \sum_{n=1}^\infty b_{nk}^\dagger w_{nk}(x) \Big\} $$
where $\{ u_{nk} \}$ and $\{ w_{nk} \}$ represent positive and negative
energy solutions.  The charge is
$$\eqalign{
Q &= \int d^2 x \, {q\over 2} \, [ \psi^\dagger, \psi ] \cr
&= q  \bigg\{ \sum_k (a_{0k}^\dagger a_{0k}^{} - {1\over 2} )
  +  \sum_{n=1}^\infty \sum_k (a_{nk}^\dagger a_{nk}^{}
  - b_{nk}^\dagger b_{nk}^{} ) \bigg\} ~~. \cr} \eqno(3)$$
When the mass $m_a$  continuously varies from positive to negative,
$u_{0k}$ becomes a negative frequency solution so that an annihilation
operator $a_{0k}$ needs to be re-interpreted as a creation operator
$b^\dagger_{0k}$ of an anti-particle.  In the expression for $Q$,
$a_{0k}^\dagger a_{0k}^{} - {1\over 2}$ is transformed into
$ b_{0k}^{} b_{0k}^\dagger - {1\over 2}
={1\over 2} - b_{0k}^\dagger b_{0k}^{}$.  This implies that $\nu_a\short{=}0$
(1) with $m_a\short{>}0$  becomes $\nu_a\short{=}1$ (0) with $m_a\short{<}0$.

Since fermions with $(\eta_a, m_a \short{<}0)$ are equivalent to those with
$(-\eta_a, |m_a| )$, a continuous change $m_a \go - m_a$ results in the
transformation of  $(\eta_a, \nu_a, |m_a|) \go (-\eta_a, 1-\nu_a, |m_a|)$.
Any physical quantities, $R$, must satisfy
$$R(\eta_a, \nu_a, m_a^2) = R( -\eta_a, 1 -\nu_a, m_a^2) ~~.
    \eqno(4)  $$

The charge  in the presence of a magnetic
field is found from (3).  For $m_a>0$
$$\eqalign{
Q_a &= q_a \, \eta_a \ep(B) \sum_k (\nu_a - \hbox{${1\over 2}$} ) \cr
&= q_a \, \eta_a \ep(B) \cdot {q_a |B|\over 2\pi}  (volume)
   \cdot (\nu_a - \hbox{${1\over 2}$} )   \cr}   $$
where we have recovered the factor $\eta_a \ep(B)$.   Hence the charge density
is found to be
$$\la j^0 (x) \ra = {1\over 2\pi} \sum_a  \eta_a q_a^2 \,
   (\nu_a - \hbox{${1\over 2}$}) \cdot B  ~, \eqno(5) $$
which satisfies (4).

An important corollary follows.  To satisfy the field
equation one must have $\kappa_0 B = \la j^0 \ra$.  Hence in order to have
$B\not= 0$,   the relation
$$ \kappa_0 = {1\over 2\pi} \sum_a  \eta_a q_a^2 \,
   (\nu_a - \hbox{${1\over 2}$})   \eqno(6) $$
must be satisfied.  This is a consistency condition for having $B\not= 0$.

One can integrate Dirac fields in (1) in a background field
$\bar F_{12} = - B$ with specified filling fractions $\{ \nu_a \}$.
Let us denote the fluctuation part of $A_\mu$ by $A'_\mu$.
The resulting effective Lagrangian  is summarized as
$$\eqalign{
\L_\eff[A'] = &- {1\over 2} B^2  + B F'_{12} \cr
&+ {1\over 2}\, F'_{0k} \, \ep \,  F'_{0k}
 - {1\over 2}\, F'_{12}\, \chi \, F'_{12}
- {1 \over 2} \, \eps A'_\mu  \, \N \,\d_\nu A'_\rho   + {\rm O}(A'^3) \cr}
   \eqno(7) $$
where
$$\eqalign{
\ep \, &= ~ 1 + \Pi_0(p) \cr
\chi &= ~ 1 + \Pi_2(p)   \cr
\kappa &= \kappa_0  - \Pi_1(p) \hskip 1cm (p_\mu\short{=}i\d_\mu) ~. \cr}
           \eqno(8)  $$
Here $\Pi_0(p)$, $\Pi_1(p)$, and $\Pi_2(p)$
represent  fermion one-loop  effects, which depend on $B$ and $\{ \nu_a
\}$.  In particular,  $\Pi_1(0)=(2\pi)^{-1} \sum_a  \eta_a q_a^2 \,
   (\nu_a - \hbox{${1\over 2}$})$.   $-\Pi_1(0)$ is the induced Chern-Simons
coefficient.   In the perturbative vacuum\myref{4}
$\Pi_1(0)_\pert  \short{=} -\sum_a \eta_a q_a^2 / 4\pi$.  The consistency
condition (6) can be written as
$$ B \cdot \kappa(0) =0 ~~.  \eqno(9) $$
To have $B\short{\not=}0$, the total Chern-Simons
coefficient at zero momentum must vanish.

\mysec{3. Energy Density}
Whether or not a non-vanishing $B$ is dynamically generated is determined
by examining the difference in the energy densities of the variational ground
state and perturbative vacuum, $\Delta \E \short{=}
\E_\gr (B ,\{ \nu_a \}) \short{-} \E_\pert$.  It is found to be
$$\eqalign{
 \Delta \E  &= ~{1\over 2} \, B^2 +  \Delta \E_{\rm f} \cr
\noalign{\kern 4pt}
&+  \int_0^1 {d\alpha\over\alpha}  ~~
i \int {d^3p\over (2\pi)^3}
{}~ \tr D_0^{-1} (p)
  \Big\{ D_\gr(p~; B,\{ \nu_a \} ) -  D_\pert(p) \Big\}  \cr}
    \eqno(10) $$
The second term, $\Delta \E_{\rm f}$, represents the shift in fermion
zero-point energies.
In the last term we have introduced an auxiriary parameter $\alpha$
representing the coupling of a fluctuation part $A_\mu'$ of gauge fields to
fermions. Full gauge field propagators $D_\gr(p)$ and $ D_\pert(p)$ are
determined with a given $\alpha$.  The formula (10) is exact, involving
no approximation.

We evaluate the gauge field propagators  in the random phase approximation
in which an infinite series of ring diagrams are summed.  It is
equivalent to keep terms in (6) up to O($A'^2$) to determine the
propagators.   Then the $\alpha$-integration in (10) can be easily
performed.   The result is
$$\eqalign{
&\Delta \E = {1\over 2} B^2 + \Delta \E_{\rm f}
 - {i\over 2} \int {d^3p\over (2\pi)^3} ~
\ln \, {\Spect(p)_\gr \over \Spect(p)_\pert }  \cr
\noalign{\kern 4pt}
&\Spect (p) =\ep^2 p_0^2 - \ep \chi {\vec p\,}^2 - \N^2  ~. \cr
\noalign{\kern 4pt}
}    \eqno(11) $$

Let us consider a chirally symmetric model consisting of
$N_{\rm f}$ pairs of $\eta_a\short{=} +$ and $-$ fermions with the same mass
$m_a\short{\ge} 0$ and  charge $q_a\short{>}0$.   In this model one has
$\sum_a \eta_a q_a^2\short{=} 0$, and
 $\Pi_1(p)_\pert\short{=}0$ exactly in the perturbative vacuum.
We suppose that the bare Chern-Simons
coefficient and  filling factors $\{ \nu_a \}$ of the variational ground state
satisfy the condition (6): $\N_0 \short{=} \sum_a \eta_a \nu_a q_a^2/2\pi
\short{\not=}0$.

In the massless fermion limit
$$\eqalign{
\noalign{\kern 4pt}
\Delta \E= & ~ {1\over 2} B^2 +
\sum_a {\zeta({3\over 2}) \over 2^{5/2} \pi^2} \, |q_a B |^{3/2} \cr
\noalign{\kern 4pt}
 &- {\sum \eta_a \nu_a q_a^3 \over 2\pi^3}  \cdot
\tan^{-1} {8 \sum \eta_a \nu_a q_a^2 \over \pi \sum q_a^2}
  \cdot |B|   + \O(|B|^{3/2}).  \cr
\noalign{\kern 4pt}
}  \eqno(12)  $$
A wide class of models give a negative coefficient for the linear term
($\short{\propto}|B|$).    As an example, suppose that $\nu_a \short{=}1$
($\nu_a \short{=}0$) for $\eta_a \short{=} +$ ($\eta_a \short{=}-$), and
$\N_0 \short{=} \sum_{\eta=+} q_a^2/2\pi$. In this case (12) becomes
$$
\Delta \E  =
- {\sum  q_a^3 \over 4\pi^3}  \cdot
\tan^{-1} {4  \over \pi }
  \cdot |B|   + \O(|B|^{3/2}) ~. \eqno(13)  $$
Hence the energy density is minimized at $B \short{\not=}0$ so that the
Lorentz invariance is spontaneously broken.\myref{5}

\mysec{4. Zero-Point Energies}
There is a simple way to understand how the linear term ($\short{\propto}
|B|$) appears in the energy density (12) and why its coefficient turns out
negative.\myref{2}   The relevant ingredient is the shift in zero-point
energies of photons.

In perturbation theory a photon is topologically massive, with a mass
given by the bare Chern-Simons coefficient $|\kappa_0|$.   In our ground
state $\Psi_\gr (B,\{ \nu_a\})$ the photon spectrum is determined by
$$\Spect (p) = \ep^2 p_0^2 - \ep \chi {\vec p\,}^2 - \N^2 =0  ~~.
          \eqno(14)   $$
The gapful or gapless nature of the spectrum is controled by the total
Chern-Simons coefficient $\kappa(p)$.   We have seen above that
$\kappa(0) \short{=}0$ in $\Psi_\gr$.  In other words, a photon becomes
gapless.

If $\kappa_0 \short{\not=}0$, the photon spectrum changes significantly,
which causes the change in zero-point energies.   It is estimated
in the following manner.

In $\Psi_\gr$, $\kappa(p)$ vanishes at $p\short{=}0$ but approaches the
perturbative value $\kappa_0$ at large $p$.  The crossover is found to take
place at a scale of the ``average'' inverse magnetic length $\ilave$:
$${1\over \slave}
   = \bigg| {\sum \eta_a \nu_a q_a^2 / l_a^2
         \over \sum \eta_a \nu_a q_a^2}  \bigg|
 = {|\sum \eta_a \nu_a q_a^3 | \over |\sum \eta_a \nu_a q_a^2|} \cdot |B|
         ~.     \eqno(15)  $$

If one suppresses the effect of $\ep(p)$ and $\chi(p)$, the photon spectrum
is approximately given by
$p_0 \short{\sim}  ({\vec p\,}^2 + \kappa^2)^{1/2}$.
Hence the shift in zero-point energies for small $|B|$ is estimated as
$$\eqalign{
\Delta \E_{\rm z.p.} &\sim \int^{\ilave}
 {d\vp\over (2\pi)^2} ~
{1\over 2} \bigg\{ \sqrt{ {\vp\,}^2} - \sqrt{ {\vp\,}^2 + \N_0^2} \bigg\}
\cr
\noalign{\kern 5pt}
&= -{1\over 8\pi} {|\N_0|\over \slave}
    + \O \Big( {1\over l_{\rm ave}^{\,3}} \Big)  \cr
\noalign{\kern 5pt}
&=  - {|\sum \eta_a \nu_a q_a^3 |\over 16\pi^2} \cdot |B| + \O (|B|^{3/2}) ~.
  \cr}  \eqno(16) $$
Comparing (12) and (16), one finds that in
a typical model the shift in zero-point energies of photons explains
about 50\% of the effect.

The crucial fact is that a photon is originally gapful, but becomes gapless
in the new ground state.  The linearity in $|B|$ is special to two spatial
dimensions.

\mysec{5. Nambu-Goldstone Theorem}

The nonvanishing $|B|$ implies that the Lorentz invariance is spontaneously
broken, which leads to  the existence of
the Nambu-Goldstone boson.  The associated Ward identities are
$$\lim_{p \go 0} p_\rho \, {\rm FT} \,
 \la {\rm T}[{\cal M}^{0j\rho} F_{0k} ] \ra =  - \ep^{jk} \la F_{12} \ra
= \ep^{jk} B ~~,  \eqno(17)  $$
where ${\cal M}^{\mu\nu\rho}(x)$ is the angular momentum density and FT stands
for a Fourier transform.  We have made use of
$\d_\sigma \la F_{\mu\nu}(x) \ra \short{=}0$.
Nonvanishing $B$ on the r.h.s.\ implies
that there must be gapless poles in the correlation function
$\la {\cal M} F \ra$ on the l.h.s..

Since we have two identities, namely for $j=1,2$, naively one
might expect two Nambu-Goldstone bosons, corresponding to two broken
Lorentz-boost generators.   However,  there seems to appear only one
Nambu-Goldstone boson which couples to both broken generators.
A photon is the Nambu-Goldstone boson, and it seems to saturate both
Ward identities.\myref{6,7}

It is straightforward to see that a photon is the Nambu-Goldstone boson
associated with the spontaneous breaking of the Lorentz invariance.   First
notice that  in terms of symmetric energy-momentum tensors
$\Theta^{\mu\nu}$, ${\cal M}^{\mu\nu\rho}\short{=} x^\mu \Theta^{\nu\rho}
-x^\nu \Theta^{\mu\rho}$.  The Ward identities (17) become
$$\lim_{p \go 0} ip_\rho \bigg\{
 {\d\over \d p_0}\, {\rm FT} \, \la {\rm T}[\Theta^{j\rho} F_{0k} ] \ra
- {\d\over \d p_j}\, {\rm FT} \, \la {\rm T}[\Theta^{0\rho} F_{0k} ] \ra
\bigg\}
= \ep^{jk} B ~~.  \eqno(18)  $$

Further we recall that
$$\eqalign{
\Theta^{\mu\nu} =& - F^{\mu\lambda} F^\nu_{~\lambda}
+ g^{\mu\nu} \, {1\over 4} F_{\rho\sigma} F^{\rho\sigma} \cr
& + \sum_a
{i\over 4} \Big\{
\psibar_a ( \gamma^\mu D^\nu + \gamma^\nu D^\mu) \psi_a
- \psibar_a (\gamma^\mu \Dnu + \gamma^\nu \Dmu ) \psi_a \Big\} \cr}
    \eqno(19) $$
In $\Psi_\gr$, $\la F_{12}\ra \short{=}-B$ and
$\la j^0 \ra \short{\propto}B$.
Hence $\Theta^{\mu\nu}$ yields terms of the form
$B \cdot F'_{\rho\sigma}$ or $B \cdot j^\rho$,  which give non-vanishing
contributions in (18).  A photon couples to both $F'^{\rho\sigma}$ and
$j^\rho$.  It is yet to be seen if the photon pole saturates the Ward
identities (18).

As mentioned in the previous section,
the gapless nature of a photon also follows from the consistency
condition (9), or $\kappa(0) \short{=} 0$.  The condition $\kappa(0)
\short{=} 0$ is vital to have non-vanishing contributions on the l.h.s.\
of (18) in the $p \short{\go} 0$ limit.

We conclude that a photon is the
Nambu-Goldstone boson associated with the spontaneous breakdown of the
Lorentz invariance.

\mysec{6. Mechanism At Work}

Now one can understand why and how the spontaneous breakdown of the Lorentz
invariance takes place in the model.  We assume $\N_0 \short{\not=}0$ so
that a photon is originally topologically massive.

Suppose that $B\short{\not=}0$ is dynamically generated.  Then it implies
that the Lorentz symmetry is spontaneously broken.  As shown above, a
photon is the  Nambu-Goldstone boson.  It becomes gapless.  This change of
the spectrum causes a shift in zero-point energies, lowering the energy
density as seen in sections 3 and 4.  Hence a non-vanishing $B$ is
prefered energetically.  The initial ansatz is justified.
The mechanism works in a self-consistent cycle.

With this peception one recognizes that the consisitency condition (6) or
(9) is viewed practically as a sufficient condition for the spontaneous
breaking of the Lorentz invariance.    If $\kappa_0 \short{\not=}0$ and
a set $\{ \nu_a \}$ satisfying (6) can be found, then $B \short{\not=}0$
is consistently realized with a lower energy density.

One can relax the condition and consider
more general models.   It is interesting to investigate chirally
asymmetric models,  models with heavy fermions, or models with
a general value of $\N_0$.   How about extensions to higher dimensions?
Can the mechanism induce spontaneous compactification in higher
dimensional theories such as string theory?   Many problems are left
for further study.

\mysec{\ninebf Acknowledgements}
This work was supported in part
by the U.S.\ Department of Energy under contract no. DE-AC02-83ER-40823.

\bigskip

\def\ap#1#2#3{{\nineit Ann.\ Phys.\ (N.Y.)} {\ninebf {#1}}, #3 (19{#2})}

\def\ijmpA#1#2#3{{\nineit Int.\ J.\ Mod.\ Phys.} {\ninebf {A#1}}, #3 (19{#2})}

\def\plB#1#2#3{{\nineit Phys.\ Lett.} {\ninebf {#1}B}, #3 (19{#2})}

\def\prl#1#2#3{{\nineit Phys.\ Rev.\ Lett.} {\ninebf #1}, #3 (19{#2})}

\def\prD#1#2#3{{\nineit Phys.\ Rev.} {\ninebf D{#1}}, #3 (19{#2})}

\def\cline{\hfil\noexpand\break  ^^J}

\def\myno#1{\item{#1.}}

\mysec{References}


\parindent=20pt
\ninerm

\myno{1}  Y.\ Hosotani, \plB {319} {93} {332}.

\myno{2}  Y.\ Hosotani, {\nineit ``Spontaneous breakdown of the Lorentz
invariance''},  UMN-TH-1238/94.

\myno{3}  See, for a different type of variational ground states,  P.\ Cea,
\prD {32} {85} {2785}.

\myno{4}  A.N.\ Redlich, \prl {52} {84} {18}; \prD {29} {84} {2366};
          K. Ishikawa, \prl {53} {84} {1615}; \prD {31} {85} {1432};
          V.Y.\ Zeitlin, {\nineit Yad.\ Fiz.} {\ninebf 49}, 742 (1989)
          [{\nineit Sov.\ J.\ Nucl.\ Phys.} {\ninebf 49}, 461 (1989)].

\myno{5}   Our picture   differs from that of  J.D.\ Bjorken, \ap {24} {63}
{174},  and that of Y.\ Nambu, {\nineit  Supplement of Prog.\ Theoret.\
Phys.} {\ninebf Extra Number}, 190 (1968).   In our case the Lorentz
invariance is  spontaneously broken physically, i.e.\ the physical content of
the spectrum is not Lorentz invariant.

\myno{6}  Y.\ Hosotani, {\nineit ``Spontaneous breakdown of the Lorentz
invariance and the Nambu-Goldstone theorem''},  UMN-TH-1304/94.

\myno{7}   A.\ Kovner and B.\ Rosenstein have attempted to regard photons
as the Nambu-Goldstone bosons associated with the spontaneous breaking of
flux symmetry in three-dimensional QED.   The Lorentz invariance is
unbroken.   A.\ Kovner and B.\ Rosenstein, \ijmpA {30} {92} {7419}.

\vfil

\end